\documentclass[10pt, aps,prl,twocolumn,showpacs,superscriptaddress,nofootinbib,preprintnumbers]{revtex4-2}

\usepackage{amsmath, graphicx}
\usepackage{xcolor}
\usepackage[colorlinks,linkcolor=blue,anchorcolor=blue,citecolor=blue,urlcolor=blue,]{hyperref}
\usepackage{orcidlink}
\usepackage{booktabs}
\usepackage{lineno}
\usepackage{float}
\usepackage{algorithm}
\usepackage{algpseudocode}
\usepackage{multirow}
\usepackage{diagbox}
\begin{document}

\title{Nonlinear growth and amplification of phase-transition gravitational waves induced by cosmic expansion}




\affiliation{School of Physics and Astronomy, Monash University, Melbourne 3800 Victoria, Australia}
\affiliation{School of Physics, Anhui University, 111 Jiulong Road, Hefei, Anhui, China 230601}
\author{Xiao Wang\,\orcidlink{0000-0003-2271-1340}}
\affiliation{School of Physics and Astronomy, Monash University, Melbourne 3800 Victoria, Australia}

\author{Chi Tian\,\orcidlink{0000-0002-5891-8573}}
\email[Contact Author:~]{ctian@ahu.edu.cn}
\affiliation{School of Physics, Anhui University, 111 Jiulong Road, Hefei, Anhui, China 230601}

\author{Csaba Bal\'azs\,\orcidlink{0000-0001-7154-1726}}
\affiliation{School of Physics and Astronomy, Monash University, Melbourne 3800 Victoria, Australia}

\date{\today}

\begin{abstract}
We perform the first three-dimensional hydrodynamical simulations of cosmological first-order phase transitions in an expanding background. These simulations consistently incorporate the effects of the evolving phase transition strength throughout the full nucleation process of slow phase transitions.
We find that, in addition to reducing mean bubble separations via an effectively enhanced nucleation rate, cosmic expansion unexpectedly induces highly nonlinear growth in the gravitational wave energy fraction, ultimately leading to a significant $\mathcal{O}(10)$ to $\mathcal{O}(100)$ amplification of the gravitational wave spectra.
This amplification is more pronounced for initially weak transitions than for those of initially intermediate strength.
Our results highlight the challenge and importance of accurately modelling slow phase transitions while accounting for cosmic expansion. 
\end{abstract}


\maketitle

\noindent\textbf{\textit{Introduction}} -- 
Gravitational waves (GWs) sourced by cosmological first-order phase transitions (FOPT) have attracted considerable attention since the first detection of GW signals announced by LIGO and Virgo~\cite{LIGOScientific:2016aoc}. 
Recent pulsar-timing array (PTA) observations by NanoGrav~\cite{NANOGrav:2023gor}, CPTA~\cite{Xu:2023wog}, EPTA~\cite{EPTA:2023fyk} and PPTA~\cite{Reardon:2023gzh} suggest the possible existence of a stochastic gravitational wave background (SGWB) peaked at  nano-Hertz frequencies. 
Cosmological FOPTs occurring at the QCD scale could provide a potential explanation for these signals. 
Furthermore, GWs generated by electroweak-scale FOPTs in the early Universe are among the primary targets of upcoming observatories, including LISA~\cite{LISA:2017pwj,LISA:2024hlh}, Taiji~\cite{Hu:2017mde}, and TianQin~\cite{TianQin:2015yph}, etc. 
Therefore, obtaining accurate theoretical predictions for the GWs sourced by FOPTs is an essential first step toward using GWs as a novel probe of physics beyond the Standard Model.

For thermal FOPTs without significant supercooling, sound waves provide the dominant contribution to the SGWB, and this component has been extensively studied using both  numerical~\cite{Hindmarsh:2013xza,Hindmarsh:2015qta,Hindmarsh:2017gnf,Cutting:2019zws,Jinno:2020eqg,Jinno:2022mie,Jinno:2024nwb,Wang:2024slx,Tian:2024ysd,Caprini:2024gyk} and analytical approaches~\cite{Hindmarsh:2016lnk,Hindmarsh:2019phv,Guo:2020grp,Wang:2021dwl,Cai:2023guc,RoperPol:2023dzg,Giombi:2024kju,Tian:2025zlo,Giombi:2025tkv}.
However, conventional methods usually neglect a crucial ingredient in the calculation: cosmic expansion, which should play an important role during both the bubble nucleation processes and subsequent GW production, especially for slow FOPTs.

Existing studies~\cite{Guo:2020grp,RoperPol:2023dzg,Xiao:2024rsj} based on the sound-shell model~\cite{Hindmarsh:2016lnk,Hindmarsh:2019phv} have attempted to incorporate cosmic expansion, generally predicting a suppressed GW signal with a shifted peak frequency. However, these semi-analytical approaches rely on assumptions such as the linear superposition of sound waves, rendering them increasingly unreliable for stronger transitions. 
Furthermore, as the plasma energy density is diluted by cosmic expansion, the phase transition strength is consequently amplified~\cite{Ellis:2019oqb,Wang:2020jrd,Athron:2023rfq}. This implies that the efficiency of energy conversion from latent heat into the fluid shells of the expanding bubbles is expected to increase.
This effect has been largely overlooked in previous studies and could introduce corrections in the prediction of GWs sourced by sound waves.

In this Letter, we present the first systematic study employing three-dimensional hydrodynamical simulations within the Higgsless framework~\cite{Jinno:2022mie} in a Friedmann–Lemaître–Robertson–Walker (FLRW)  background, incorporating a time dependent phase-transition strength. We find that cosmic expansion effectively enhances the nucleation rate, thereby reducing the mean bubble separation and consequently shifting the peak frequency to higher values. Remarkably, cosmic expansion further induces nonlinear growth in the GW energy fraction, ultimately leading to a substantial amplification of the GW signal. Moreover, depending on the sound-wave lifetime, this amplification can be stronger for initially weak FOPTs than for those of initially intermediate strength.

\noindent\textbf{\textit{Dynamics in an expanding Universe}} -- 
We model the fluid during the phase transition using the bag equation of state (EoS).
The energy density $\rho$ and pressure $p$ are given by
\begin{equation}
    \rho = a_*T^4 + \epsilon, \quad p=\frac{1}{3}a_*T^4 - \epsilon, 
\end{equation}
where $\epsilon=0$ in the broken phase and $\epsilon>0$ in the symmetric phase, and $a_*=g_*\pi^2/30$, with $g_*$ denoting the number of relativistic degrees of freedom.
We can therefore define the strength parameter as 
\begin{equation}
    \alpha = \frac{\epsilon}{a_*T^4}=\frac{4}{3}\frac{\epsilon}{w}\,,
    \label{eq:alpha}
\end{equation}
where $w=\rho+p$ is the enthalpy.
Since we focus on GWs sourced by sound waves, the energy momentum tensor (EMT) in the expanding Universe, neglecting the scalar field contribution, is given by 
\begin{equation}
    T^{\mu\nu} = (\rho + p)U^\mu U^\nu + g^{\mu\nu}p\,,
\end{equation}
where the four-velocity is defined in a flat FLRW  spacetime, with $ds^2 = - dt^2 + a^2(t)(dx^2 + dy^2 + dz^2)$, is defined as $U^\mu = \gamma(1 , v_x/a(t), v_y/a(t), v_z/a(t))$
with $\gamma\equiv1/\sqrt{1 - v^iv_i}$ and $v^i = d x^i / d \eta$, where $\eta$ is the conformal time.
The evolution of the scale factor $a(t)$ follows the Friedmann equation $H^2\equiv[\dot{a}(t)/a(t)]^2=\rho_{\rm tot}/(3m_p^2)$, where the overdot denotes differentiation with respect to physical time. The total density $\rho_{\rm tot}$ encompasses contributions from both radiation and vacuum energy, and $m_p$ denotes the reduced Planck mass. In this study, we restrict our analysis to FOPTs with weak or intermediate strength, where $\alpha \ll 1$ and radiation dominates. Consequently, when solving the Friedmann equation in our study, we adopt the approximation $\rho_{\rm tot} \simeq \rho_{\rm rad}$.

\begin{figure}[t!]
    \centering
    \includegraphics[width=0.492\linewidth]{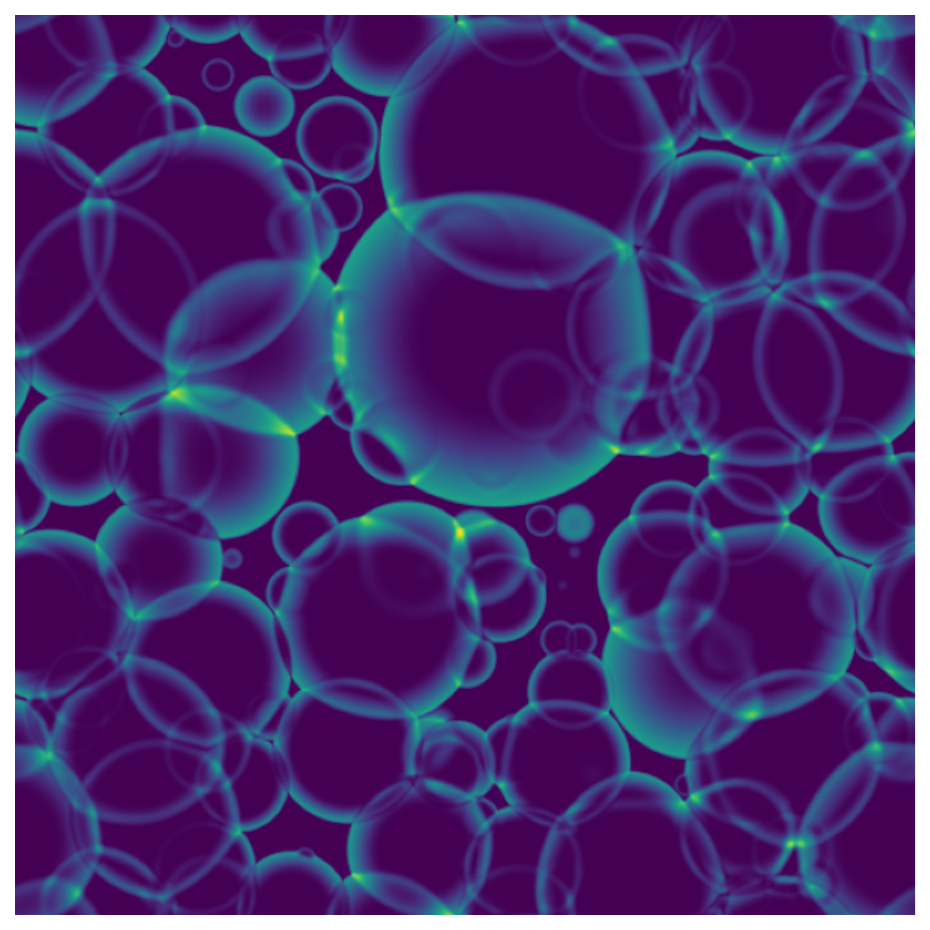}
    \includegraphics[width=0.492\linewidth]{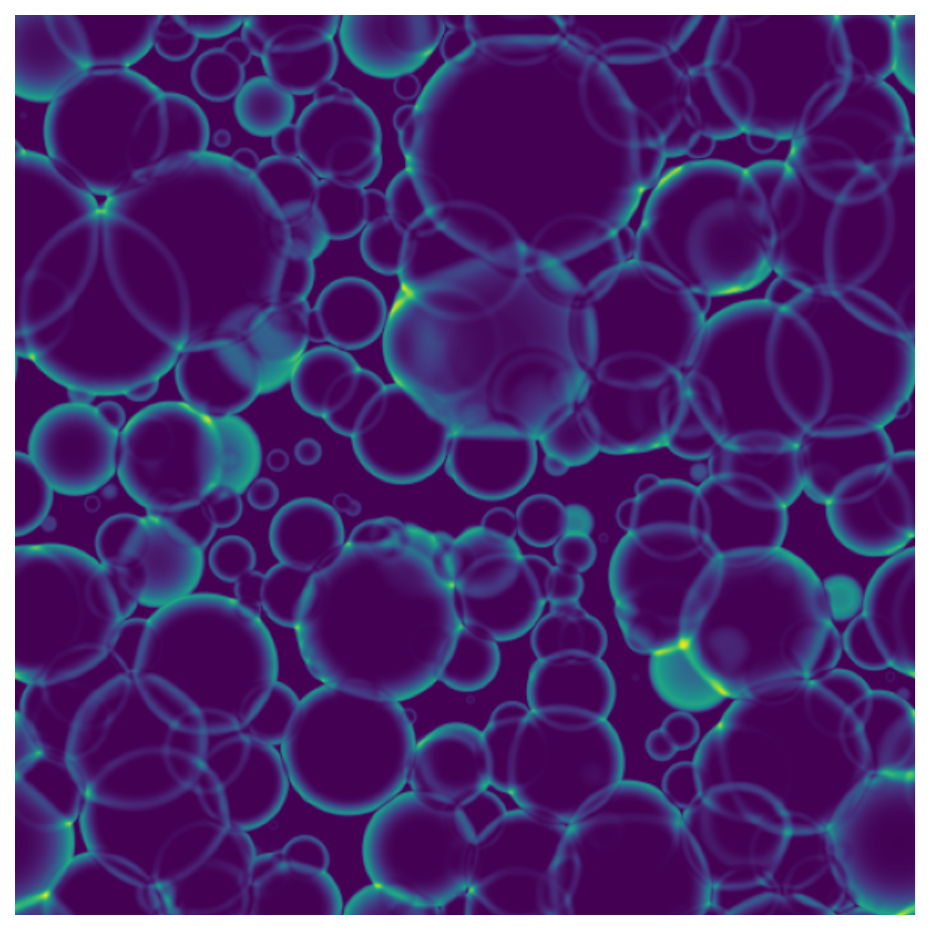}
    \caption{Snapshots of the  kinetic energy density $w\gamma v^2$ around the percolation time $t\approx 10 \beta^{-1}$. The left panel depicts the slice without expansion, whereas the right panel shows the slice with cosmic expansion.}
    \label{fig:slices}
\end{figure}

When cosmic expansion is taken into account, an implicit effect arises that has been overlooked in most previous studies: the strength parameter is inherently time dependent. This occurs because the radiation energy density decreases rapidly as the Universe expands, whereas the vacuum energy density remains unchanged. According to the definition in Eq.~\eqref{eq:alpha}, this generally results in a time dependent $\alpha(t)$ that increases monotonically with time.
Previous studies~\cite{Espinosa:2010hh,Giese:2020rtr,Giese:2020znk,Wang:2020nzm,Wang:2023jto} have shown that a larger strength parameter $\alpha$ generally enhances the kinetic energy fraction. Because the strength of the GWs sourced by sound waves is proportional to this kinetic energy fraction~\cite{Caprini:2019egz}, a larger $\alpha$ typically yields a stronger GW signal.

To make precise predictions of GW signals under cosmic expansion, we present a methodology incorporating the FLRW background into full 3D hydrodynamical simulations. 
We generalise the strategy adopted in spherically symmetric simulations~\cite{Jinno:2024nwb} and factorise the thermodynamic quantities as $X(t, x)\equiv X_{\rm rel}(t, x) X_b(t)$, where the homogeneous background $X_b(t)$ accounts for the effect of cosmic expansion.
In addition, we normalise the conservation equations $\nabla_\mu T^{\mu\nu}=0$ with background enthalpy $w_b(t)$, and denote the normalised quantities by $\tilde{X}=X/w_b$, where the background quantities satisfy $\rho_b = \frac{3}{4}w_b + \epsilon$ and  $p_b = \frac{1}{4}w_b - \epsilon$ as the bag model is adopted.
Using the continuity equation of the background quantities $\dot{\rho}_b + 3H(\rho_b + p_b) = 0$, we obtain $w_b(t)\propto a(t)^{-4}$.
Thus, the effective strength parameter grows as the fourth power of the scale factor: $\alpha(t) = 4\epsilon/(3w_b(t)) \equiv \alpha_0 a(t)^4$, where the subscript $0$ denotes the initial value at the onset of the phase transition.
The conservation of the EMT in the expanding Universe then yields
\begin{align}
\partial_t{\tilde{E}} + \frac{1}{a}\partial_i \tilde{Z}^i - H(\tilde{E} - 3\tilde{p} - \tilde{Z}_i v^i) = 0\label{eq:nmfleqv21}\,,\\
\partial_t \tilde{Z}_i + \frac{1}{a}\partial_j (\tilde{Z}_iv^j) + \frac{1}{a}\partial_i\tilde{p} = 0\,,\label{eq:nmfleqv22}
\end{align}
where $\tilde{E} = (w\gamma^2 - p)/w_b$ and $\tilde{Z}_i = (w\gamma^2 v_i)/w_b$.
These equations consistently incorporate both the effects of cosmic expansion and the increasing strength of the phase transition during the FOPT.
To solve the fluid equations above, we employ the Higgsless framework~\cite{Jinno:2022mie}, where the expanding bubbles are simulated by incorporating a moving boundary with a constant wall velocity $v_w$, and the bag constant $\epsilon$ differs between the interior and exterior regions of the bubbles. Furthermore, the Kurganov-Tadmor scheme~\cite{Kurganov:2000ovy} is utilised to correctly resolve the shock fronts. 
For simplicity and computational efficiency, we also neglect the last term in the bracket of Eq.~\eqref{eq:nmfleqv21}, as it has a negligible impact on the final results. Unless otherwise specified, all numerical results are reported from simulations with a resolution $N=512$.

In addition, the effects of cosmic expansion also manifest in the bubble nucleation history. The bubble nucleation rate is estimated as $ \Gamma(t) = H_n^4 \exp (\beta t)$,
where $\beta$ is the inverse duration of the phase transition, and we set the nucleation time to $t_n = 0$, corresponding to the time at which there is, on average, one bubble per Hubble volume $H_n^{-3}$. In an expanding Universe, during the nucleation process for relatively slow phase transitions with $\beta / H_n \sim 20$, the growing physical size effectively enhances the number of nucleation events per comoving volume.

The nucleation history is also closely related to the bubble wall velocity $v_w$, which is intrinsically time-dependent and can be estimated by solving the Boltzmann equations together with the scalar field equation of motion~\cite{Moore:1995si,Konstandin:2014zta,Dorsch:2021ubz,Dorsch:2021nje,Laurent:2022jrs,Wang:2020zlf,Jiang:2022btc,Wang:2024slx}, or alternatively, by employing various approximations~\cite{Ai:2023see,Ai:2024btx,Krajewski:2026kcm}. However, for simplicity, we treat $v_w$ as a constant in the present study and leave the investigation of time-dependent wall velocities for future work. 

The simulation box size is set to $L=40 v_w / \beta$ to ensure that the number of bubbles remains approximately constant across various wall velocities. The detailed procedure for simulating this stochastic process is described in the Supplemental Material. We find that, regardless of whether cosmic expansion is considered, the true vacuum comes to dominate the entire volume at approximately $t\approx10 / \beta$, which is referred to as the percolation time. However, incorporating cosmic expansion results in smaller bubble separations, causing the bubbles to appear more crowded than in the non-expanding cases, as illustrated in Fig.~\ref{fig:slices}.

\noindent\textbf{\textit{GW production}} -- 
The stress-energy tensor $T_{\mu \nu}$ raised by the dynamics of the radiation fluid excites metric perturbations. In Fourier space, the time evolution of the perturbed metric $ {h}_{ij}(t,\mathbf{k})$ reads:
\begin{align}
\label{eq:hij_dot}
    \ddot{h}_{ij}(t,\mathbf{k}) + 3H \dot{h}_{ij}(t,\mathbf{k}) + \frac{\mathbf{k}^2}{a^2} h_{ij}(t,\mathbf{k})
= \frac{2\Pi^{\rm TT}_{ij}(t,\mathbf{k})}{m_p^2}\,,
\end{align}
where $\Pi^{\rm TT}_{ij}(t,\mathbf{k})$ is the transverse-traceless part of the anisotropic stress tensor $\Pi_{ij}\equiv (T_{ij} - pg_{ij})/a^2$.
The energy density of the resulting SGWB at time $t$ per logarithmic interval can be computed by
\begin{align}
\label{eq:rho_gw}
    \frac{d\rho_{\mathrm{GW}}(t)}{d\ln k}
= \frac{m_p^2 k^3}{8\pi^2 V}
\int \frac{d\Omega_k}{4\pi}\,
\dot{h}_{ij}(\mathbf{k},t)\,\dot{h}_{ij}^*(\mathbf{k},t).
\end{align}
To facilitate numerical integration, we define $\tilde{h}_{ij}\equiv h_{ij} m_p^2 / w_b(0)$. With this redefinition, Eq.~\eqref{eq:hij_dot} for $\tilde{h}_{ij}$ effectively adopts $m_p=1$, and the stress energy tensor in Eq.~\eqref{eq:hij_dot} depends on $w_{\rm rel}(t,x)\equiv w(t,x) / w_b(t) = w(t,x) a(t)^{4}/ w_b(0)$. Then, the time dependent GW energy fraction can be computed by
\begin{equation}
\begin{split}
    \label{eq:Omega_gw}
    \Omega^*_{\rm GW}(t, k) &\equiv \frac{1}{\rho_{\rm tot}}\frac{d\rho_{\mathrm{GW}}(t)}{d\ln k}\\
&\simeq\frac{2}{3}\frac{H_n^2 a^4 k^3}{\pi^2 V}\int \frac{d\Omega_k}{4\pi}\,
\dot{\tilde{h}}_{ij}(\mathbf{k},t)\,\dot{\tilde{h}}_{ij}^*(\mathbf{k},t),
\end{split}
\end{equation}
where we have adopted the Friedmann equation $\rho_{\rm tot} = 3 H^2 m_p^2$ and the equation of state $\mathbf{\omega} \equiv w_b / \rho_{\rm tot} \simeq 4/3$.
In addition, by integrating all the wave-numbers in $\Omega^*_{\rm GW}(t, k)$ and defining $\tilde{\Omega}^*_{\rm GW}(t) \equiv \int \Omega^*_{\rm GW}(t, k) d \ln k$, we can identify the growth of the gravitational-wave content from the energy fraction at $t$.

\begin{figure}
    \centering
    \includegraphics[width=0.9\linewidth]{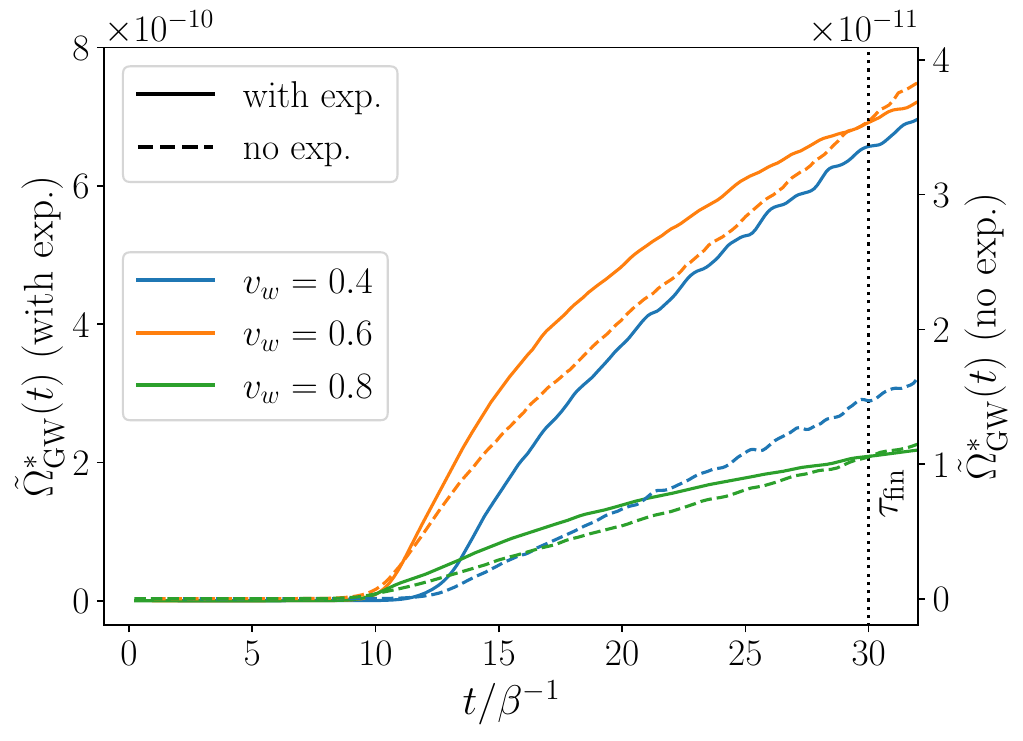}
    \includegraphics[width=0.92\linewidth]{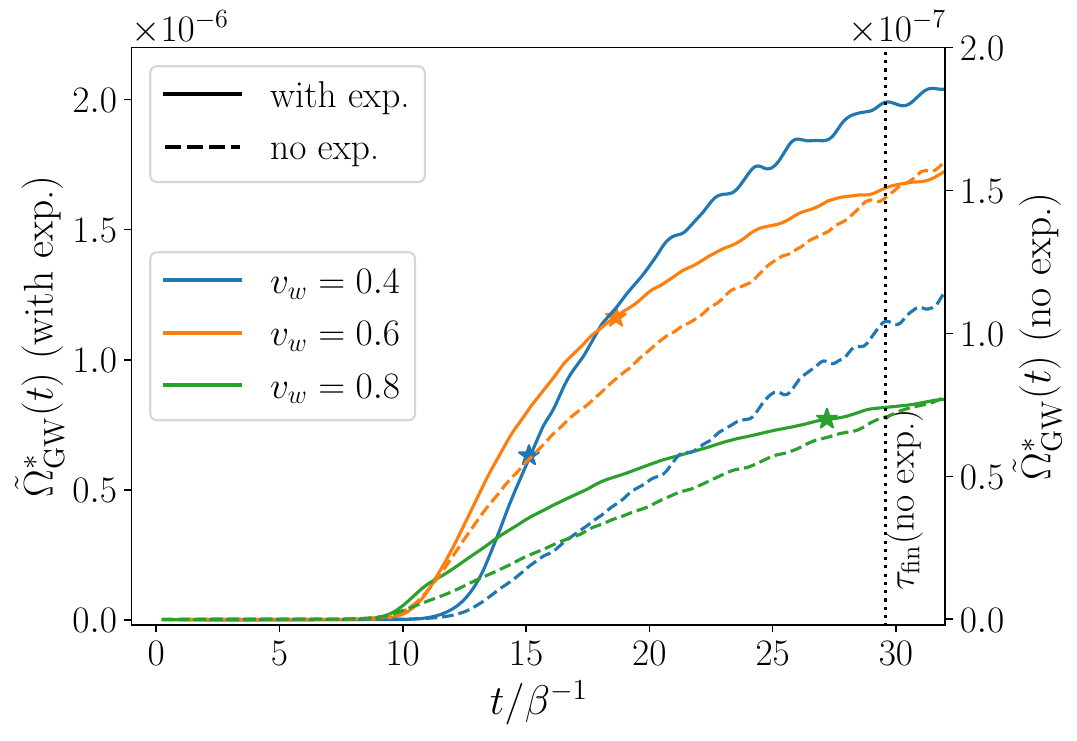}
    \caption{The time evolution of the GW energy fraction $\tilde{\Omega}_{\rm GW}^*(t)$ for various wall velocities $v_w$ with $\beta/H_n=20$. The top panel is for initially weak phase transition ($\alpha_0=0.005$) and the bottom is for intermediate phase transition ($\alpha_0=0.05$). The dashed and solid lines represent the results of simulations without and with cosmic expansion, respectively. The stars in the bottom panel represent the estimated $\tau_{\rm fin}$ when accounting for cosmic expansion.}
    \label{fig:gw_grow}
\end{figure}

In the absence of cosmic expansion, the energy of the SGWB should continue to grow as long as sound waves propagate after the completion of the FOPT. Furthermore, this growth in GW energy exhibits a linear time dependence in the non-expanding case, as supported by various numerical hydrodynamical simulations~\cite{Hindmarsh:2013xza,Cutting:2019zws,Caprini:2024gyk}.
As illustrated in Fig.~\ref{fig:gw_grow}, this linear growth is also observed in our non-expanding simulations: the $\tilde{\Omega}^*_{\rm GW}(t)$ curves remain small during the early stages of nucleation and increase linearly following percolation. This behaviour is expected, as no dissipative effects are included and sound waves persist indefinitely within the periodic simulation box.

\begin{figure*}[t!]
    \centering
    \includegraphics[width=0.32\linewidth]{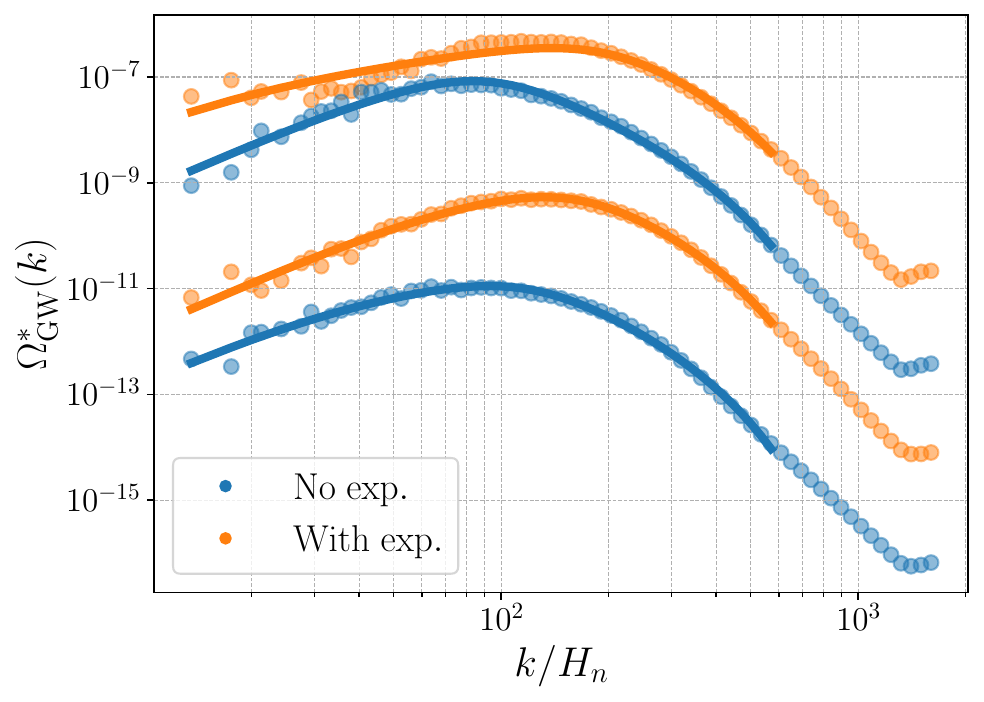}
    \includegraphics[width=0.32\linewidth]{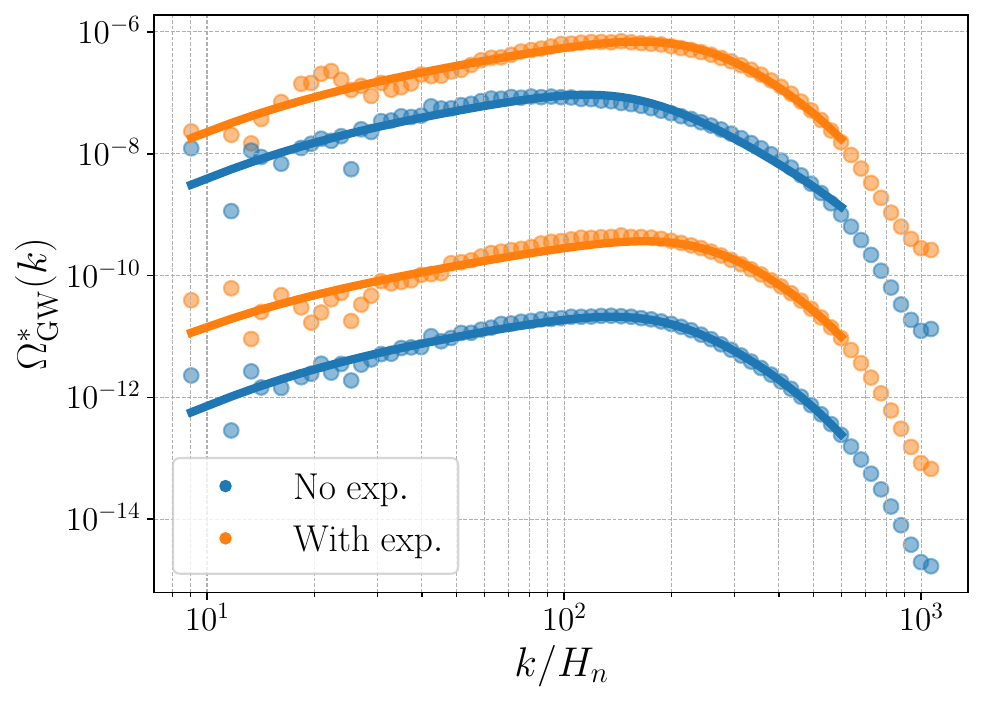}
    \includegraphics[width=0.32\linewidth]{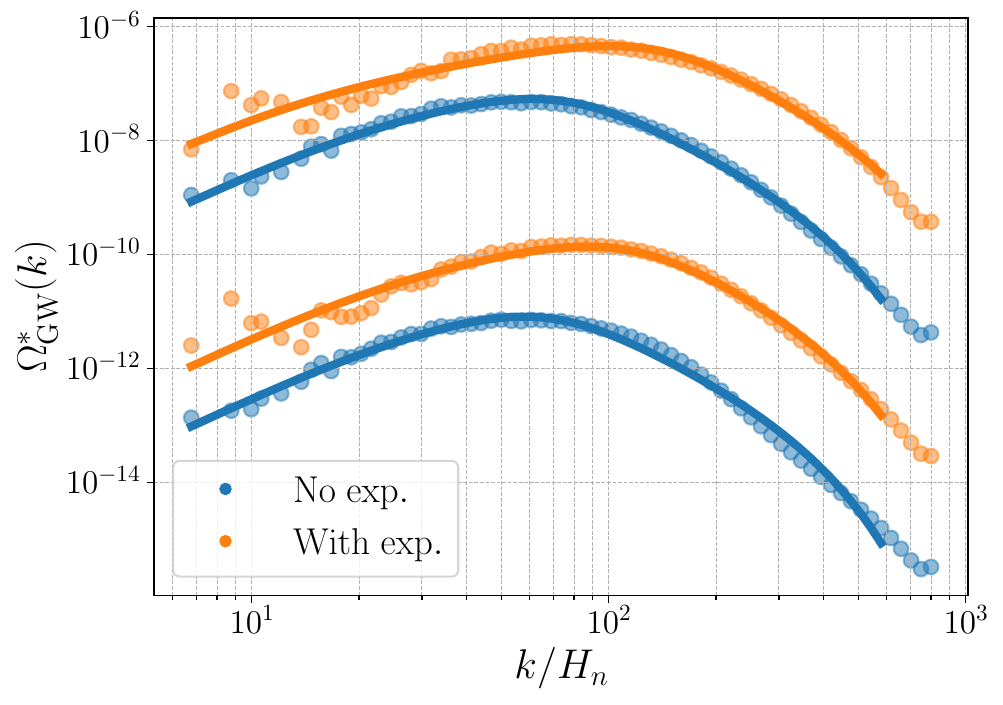}
    \caption{GW spectra at $\tau_{\rm fin}$ for various bubble wall velocity ($v_w=0.4, 0.6, 0.8$ from left to right) with $\beta/H_n=20$. The upper two curves represent the spectra of a non-expanding and an expanding universe for a phase transition with initially intermediate strength ($\alpha_0=0.05$), while the lower lines show the corresponding results of initially weak phase transitions ($\alpha_0=0.005$).}
    \label{fig:Omega_k}
\end{figure*}

\begin{figure}
    \centering
    \includegraphics[width=0.92\linewidth]{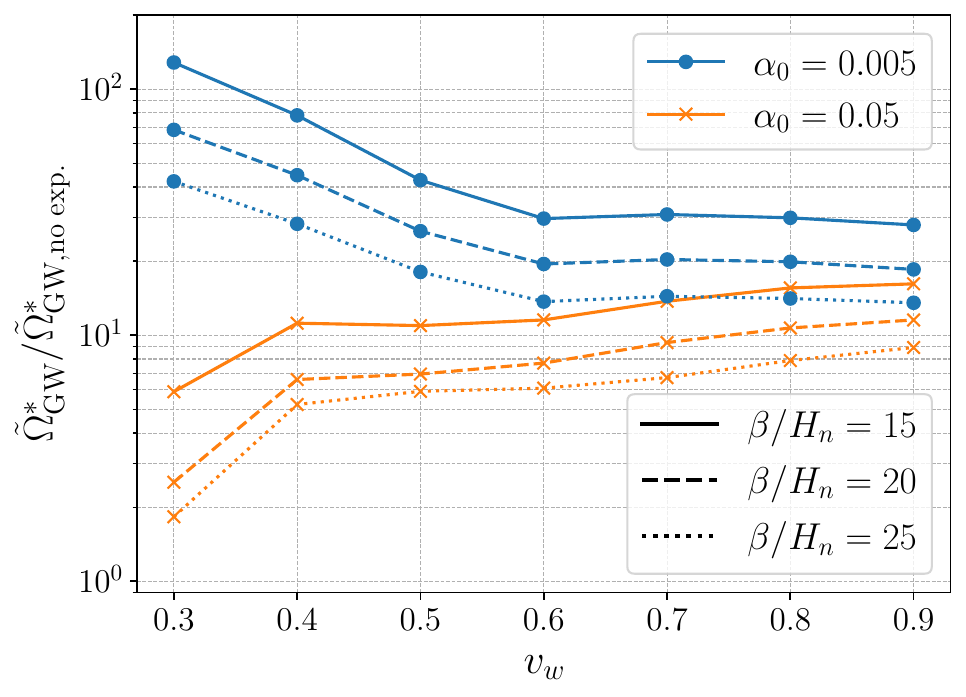}
    \caption{The ratio of $\tilde{\Omega}^*_{\rm GW}$ / $\tilde{\Omega}^*_{\rm GW, no\;exp.}$ at $\tau_{\rm fin}$ for various wall velocities with initially weak and intermediate FOPTs. }
    \label{fig:Omega_ratio}
\end{figure}

However, as also shown in Fig.~\ref{fig:gw_grow}, once cosmic expansion is taken into account, the linear growth in the GW energy breaks down.
Instead, after percolation is reached, the GW energy shows nearly exponential growth during the early stage of the growth phase, leading to a significant enhancement for various $v_w$ relative to the non-expanding case. 
This exponential growth of the GW energy fraction soon becomes suppressed once the transition is completed, resulting in a suppressed production rate of GWs, as evidenced by the decreasing growth rate of the GW energy fraction shown in Fig.~\ref{fig:gw_grow}. 
Overall, the linear growth breaks down in the presence of cosmic expansion, and the energy stored in GWs can no longer be considered strictly proportional to the sound-wave lifetime. Then, the duration of GW emission driven by sound waves, $\tau_{\rm fin}$, plays a key role in determining the final GW energy fraction.
 
An estimate of $\tau_{\rm fin}$ can be obtained by using the sound-wave lifetime $\tau_{\rm sw}$, via $\tau_{\rm fin} = \min(\tau_{\rm sw}, 1/H_n ) + \tau_{\rm poc}$, where $\tau_{\rm poc}$ denotes the completion time at which the true-vacuum bubbles nearly fill the entire Universe.
Owing to the exponential growth of the bubble nucleation rate, there is no fundamental difference in $\tau_{\rm poc}$ between the expanding and non-expanding cases; 
we therefore set $\tau_{\rm poc} = 10 / \beta$ for both scenarios. 
A typical estimation of sound-wave lifetime $\tau_{\rm sw}$ is given by the eddy turnover time~\cite{Hindmarsh:2017gnf,Caprini:2019egz}
$\tau_{\rm sw} \sim R_* / \overline{U}_f$, where $R_*$ is the bubble separation, estimated as $R_*\approx (8\pi)^{1/3} v_w / \beta$, and the RMS four-velocity is defined as $\overline{U}_f \equiv \sqrt{\frac{1}{w_b} \langle w \gamma^2 v^2 \rangle}$. 
We adopted this estimation of $\tau_{\rm sw}$ in this study, and in the scenario including cosmic expansion, $\tau_{\rm sw}$ is computed in the conformal time $\eta$.
To facilitate a direct comparison with the non-expanding case on equal footing, it is subsequently converted to the coordinate time $t$.
In practice, $\overline{U}_f$ is not exactly conserved during the simulation. 
Instead, it gradually decreases after reaching a maximum. We therefore use this maximum value to compute $\overline{U}_f$ and subsequently $\tau_{\rm sw}$. 

For initially weak FOPTs, the relatively small $\overline{U}_f$ causes $\tau_{\rm fin}$ to approach the Hubble time regardless of whether cosmic expansion is considered, as suggested by the upper panel of Fig.~\ref{fig:gw_grow}. 
Consequently, the GW energy fraction in an expanding background tends to be enhanced relative to the no-expansion case. 
This enhancement is more pronounced for smaller bubble wall velocities, corresponding to the deflagration regime.

For FOPTs with initially intermediate strength, $\tau_{\rm fin}$ becomes shorter than the Hubble time and differs between the expanding and non-expanding cases, as shown in the bottom panel of Fig.~\ref{fig:gw_grow}. This behaviour arises from the reduced mean bubble separation together with the enhanced kinetic energy fraction associated with the RMS four-velocity $\overline{U}_f$. In this case, the GW energy fraction remains higher than that in the non-expanding scenario, although the enhancement is less pronounced than for FOPTs with initially weak strength. Furthermore, as shown in the bottom panel of Fig.~\ref{fig:gw_grow}, once $\tau_{\rm fin}$ becomes shorter than a Hubble time, it decreases with a decreasing $v_w$, which further suppresses the GW energy in an expanding background for intermediate FOPTs.

The GW spectra at $\tau_{\rm fin}$ for FOPTs with initially weak and intermediate strengths are presented in Fig.~\ref{fig:Omega_k}, alongside curves fitted by a double broken power-law template, following Ref.~\cite{Jinno:2022mie}. A table containing the fitted parameters for different wall velocities and inverse durations of FOPT, $\beta/H_{n}$, is provided in the Supplemental Materials.
Compared to non-expanding scenarios, the inclusion of cosmic expansion appears to leave the spectral slopes unchanged at both low and high frequencies while shifting the peak frequency to higher values, as anticipated from the reduced mean bubble separation. Furthermore, a significant amplification of the peak amplitude, of order $\mathcal{O}(100)$, is observed for weak FOPTs. This amplification becomes less pronounced for intermediate FOPTs and further diminishes for smaller bubble wall velocities due to the decreasing $\tau_{\rm fin}$.

This order-of-magnitude amplification is generally consistent with the enhancement of $\overline{U}_f$ in the expanding scenario relative to the non-expanding case. For example, when $\alpha_0=0.005$, $v_w=0.4$, and $\beta/H_n = 20$, $\overline{U}_f$ is approximately four times larger at $\tau_{\rm fin}$ than in the non-expanding case (see Supplemental Materials for a comprehensive demonstration of $\overline{U}_f$). 
Based on previous studies~\cite{Hindmarsh:2017gnf,Cutting:2019zws}, for the non-expanding case, the GW signal is approximately proportional to $\overline{U}_f^4$, which corresponds to an expected enhancement of $\mathcal{O}(200)$. However, the final amplification is further suppressed by the reduced mean bubble separation and modified by the overall nonlinear growth, ultimately reducing the enhancement factor to $\mathcal{O}(50)$. Given that the enhancement of $\overline{U}_f$ largely stems from the amplified $\alpha(t)$, the corresponding GW enhancement is likewise attributed to the effective growth of $\alpha(t)$.

In Fig.~\ref{fig:Omega_ratio}, we present the ratio of the GW energy fraction $\tilde{\Omega}^*_{\rm GW}/\tilde{\Omega}^*_{\rm GW, no\;exp.}$ at $\tau_{\rm fin}$ for various bubble wall velocities and inverse durations in initially weak and intermediate FOPTs. Incorporating cosmic expansion evidently amplifies the GW spectra in both initially weak and intermediate FOPTs. This amplification is more pronounced for transitions with lower wall velocities and smaller inverse durations, and it decreases as $v_w$ or $\beta/H_n$ increases for initially weak transitions. As the wall velocity increases, the amplification tends to approach a constant value for a given inverse duration. 
However, this scaling relation on $v_w$ is reversed for initially intermediate FOPTs because the sound-wave lifetime is significantly suppressed at smaller $v_w$ in the expanding scenarios, as indicated by the star markers in Fig.~\ref{fig:gw_grow}. Our results suggest that different approaches to estimating the sound-wave lifetime could significantly alter the final GW spectra. Therefore, obtaining more accurate results requires a more careful determination of the sound-wave lifetime, which deserves further investigation.

\noindent\textbf{\textit{Conclusion}} --
In this Letter, we investigate the generation of GWs by slow cosmological FOPTs in an FLRW background. 
Overall, cosmic expansion is expected to modify GW production in three distinct ways. 1) A reduced mean bubble separation, as the expanding physical volume effectively enhances the nucleation rate. 2) An increasing phase transition strength due to the dilution of the radiation fluid. 3) The modification of post-transition hydrodynamics by the background expansion. Starting from simulations of the Poissonian nucleation process and employing Higgsless 3D hydrodynamical lattice simulations, our approach consistently captures all these effects within an expanding Universe.

Our results demonstrate that incorporating cosmic expansion can result in a significant amplification of the GW signals generated by slow FOPTs, contrary to previous analyses relying on semi-analytical models~\cite{Guo:2020grp,RoperPol:2023dzg,Lewicki:2025hxg}. 
This amplification stems from the nonlinear, nearly exponential growth of the GW energy density fraction during the early stage of GW production, as opposed to the linear growth observed in the non-expanding case. This rapid growth is predominantly attributed to an effective increasing phase transition strength, $\alpha(t)$, during the nucleation, an effect that cannot be adequately captured by semi-analytical frameworks. 
Although the time dependence of $\alpha(t)$ may modify the wall velocity in more realistic models, such modifications are unlikely to completely eliminate the $\mathcal{O}(100)$ amplification. 
A comprehensive assessment of the impact of a time dependent $v_w$ requires a more sophisticated treatment, which we leave for future work.


\begin{acknowledgments}
\noindent\textbf{\textit{Acknowledgments}} -- 
X.W. and C.B. are supported by Australian Research Council grants DP220100643 and LE250100010. C.T. is supported by the National Natural Science Foundation of China (Grants No. 12405048).  
\end{acknowledgments}

\bibliography{biblio.bib}

\newpage
\clearpage
\onecolumngrid

\appendix
\section*{Supplemental materials}

\subsection{Simulations of Nucleation Histories in an Expanding Universe}
\label{sm:1}

To derive the GW spectrum generated by sound waves in an expanding Universe, in addition to solving the fluid equations derived from energy-momentum conservation, simulating the bubble nucleation history simultaneously is essential. Here, we present the details of the algorithm used to simulate the bubble nucleation history in an expanding Universe.

For estimates of the nucleation rate $\Gamma(t) = H_n^4 \exp(\beta t)$ in an FLRW Universe, where the physical volume continuously expands, the increasing physical size of a comoving volume effectively enhances the nucleation rate. Therefore, we develop the following algorithm to simulate the bubble nucleation history while incorporating these modifications from cosmic expansion.
\begin{algorithm}[H]
\caption{Simulating Bubble Nucleation History in an Expanding Universe}
\label{alg:bubble_nucleation}
\begin{algorithmic}[1]
    \Require Initial time $t_i$, final time $t_f$, nucleation rate $\Gamma(t)$, volume $V(t)=V_0 a^3(t)$
    \Require Comoving box size $L$ ($V_0 = L^3$), bubble wall velocity $v_w$, conformal time $\eta(t) = \int \frac{dt'}{a(t')}$
    \Ensure Filtered list of true nucleated bubbles $\mathcal{B}_{\text{filtered}}$
    
    \State Set the current time $t \gets t_i$
    \State Initialize the raw bubble list $\mathcal{B} \gets \emptyset$
    
    \State \Comment{\textbf{Phase 1: Generate all potential bubbles}}
    \While{$t < t_f$}
        \State Generate a random number $p$ from the uniform distribution $\mathcal{U}(0, 1)$
        
        \State Solve for $t_{\text{next}}$ such that
        \begin{equation*}
            \int_{t}^{t_{\text{next}}} \Gamma(t') V(t') \, dt' = -\ln(p)
        \end{equation*}
        
        \If{$t_{\text{next}} \geq t_f$}
            \State \textbf{break}
        \EndIf
        
        \State Choose a random comoving position $\mathbf{x}$ uniformly within the comoving box $L^3$
        \State Add the bubble event $(t_{\text{next}}, \mathbf{x})$ to $\mathcal{B}$
        \State Update the current time $t \gets t_{\text{next}}$
    \EndWhile
    
    \State \Comment{\textbf{Phase 2: Filter bubbles covered by previously generated ones}}
    \State Initialize the filtered bubble list $\mathcal{B}_{\text{filtered}} \gets \emptyset$
    
    \For{\textbf{each} bubble $(t_i, \mathbf{x}_i)$ \textbf{in} $\mathcal{B}$} \Comment{List is inherently in chronological order}
        \State $is\_covered \gets \textbf{false}$
        
        \For{\textbf{each} accepted bubble $(t_j, \mathbf{x}_j)$ \textbf{in} $\mathcal{B}_{\text{filtered}}$}
            \State Calculate conformal time difference $\Delta \eta = \eta(t_i) - \eta(t_j)$
            \State Calculate comoving distance $d_c$ between $\mathbf{x}_i$ and $\mathbf{x}_j$ in the periodic box
            
            \If{$d_c / v_w < \Delta \eta$}
                \State $is\_covered \gets \textbf{true}$
                \State \textbf{break} \Comment{Bubble is covered; skip remaining checks}
            \EndIf
        \EndFor
        
        \If{\textbf{not} $is\_covered$}
            \State Add $(t_i, \mathbf{x}_i)$ to $\mathcal{B}_{\text{filtered}}$
        \EndIf
    \EndFor
    
    \State \Return $\mathcal{B}_{\text{filtered}}$
\end{algorithmic}
\end{algorithm}

By setting the scale factor $a(t)$ to unity, this procedure also applies to the case in which cosmic expansion is neglected. The numbers of bubbles generated by the above algorithm for different $\beta/H_n$ and $v_w$ in an $L=40v_w/\beta$ box are shown in Tab.~\ref{tab:bubbles_exp}. In non-expanding scenarios, the number of bubbles remains unchanged, as expected from the scaling of the box size. When cosmic expansion is included, the generated number of bubbles decreases slightly for larger $\beta$ or larger $v_w$.

\begin{table}[t!]
\centering
\begin{tabular}{c|c|c|c}
\hline\hline
$\beta/H_n$ & $v_w$ & No. bubbles without exp. & No. bubbles with exp. \\
\hline
\multirow{3}{*}{15} 
& 0.3 & 2576 & 11809 \\
& 0.6 & 2582 & 10212 \\
& 0.9 & 2600 & 9340 \\
\hline
\multirow{3}{*}{20} 
& 0.3 & 2576 & 9769 \\
& 0.6 & 2577 & 8648 \\
& 0.9 & 2582 & 7946 \\
\hline
\multirow{3}{*}{25} 
& 0.3 & 2575 & 8522 \\
& 0.6 & 2576 & 7575 \\
& 0.9 & 2577 & 7091 \\
\hline\hline
\end{tabular}
\caption{Number of bubbles with and without expansion for different $\beta/H_n$ and $v_w$ in a $L=40v_w / \beta$ box.}
\label{tab:bubbles_exp}
\end{table}

\subsection{RMS velocities and convergence test}

\begin{figure}[t!]
    \centering
    \includegraphics[width=0.5\linewidth]{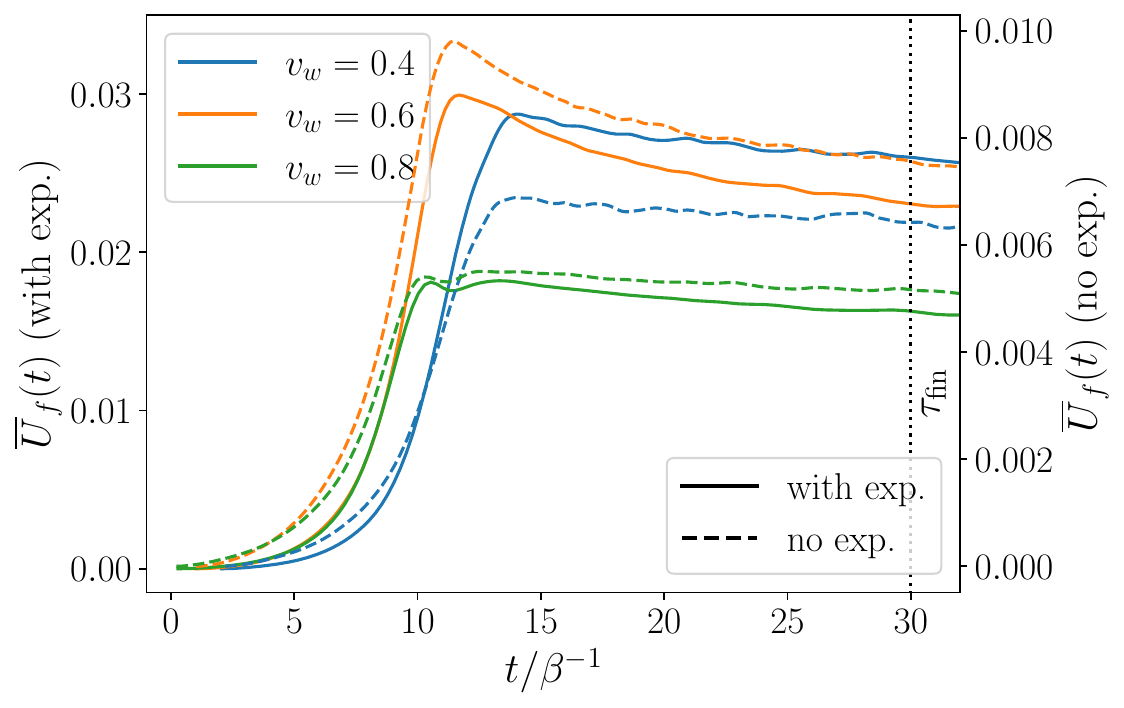}%
    \includegraphics[width=0.5\linewidth]{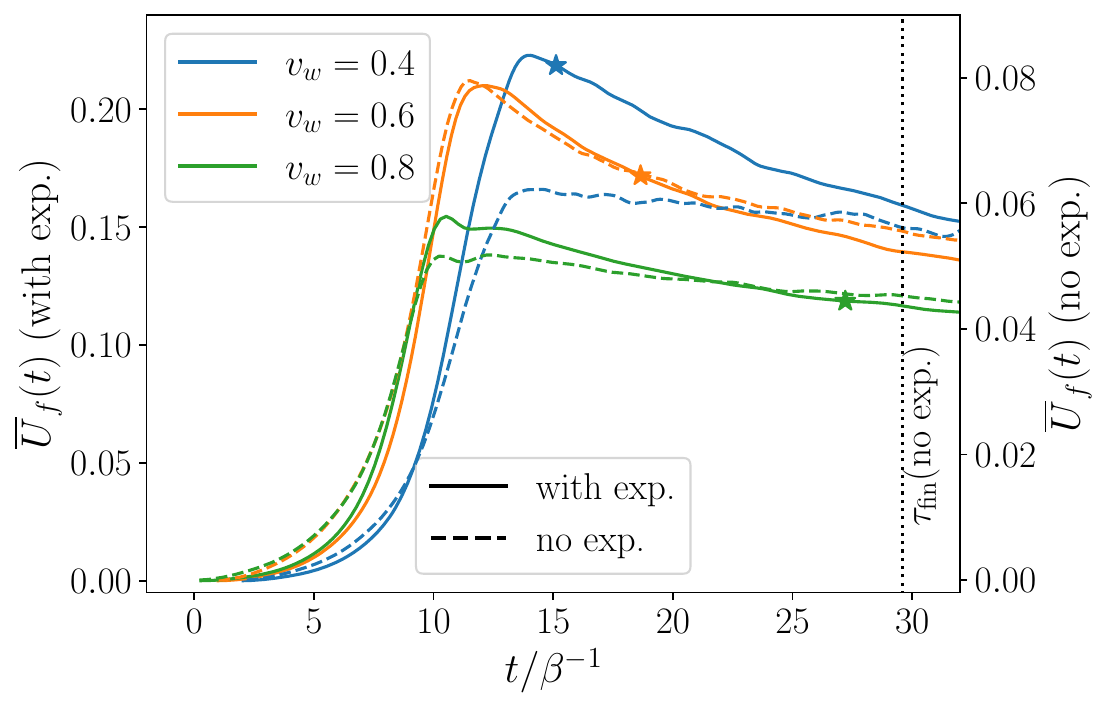}
    \caption{RMS velocities $\overline{U}_f(t)$ for various wall velocity $v_w$ with $\beta/H_n=20$. The left panel is for weak phase transition ($\alpha_0=0.005$) and the right is for intermediate phase transition ($\alpha_0=0.05$). The dashed and solid lines represent the result of simulations without and with cosmic expansion. }
    \label{fig:Uf_t}
\end{figure}

In Fig.~\ref{fig:Uf_t}, we present the time evolution of RMS velocities in the expanding and non-expanding scenarios. Although they are not exactly conserved numerically during the simulation, the decreasing trends in both cases are similar.

\begin{figure}[t!]
    \centering
    \includegraphics[width=0.5\linewidth]{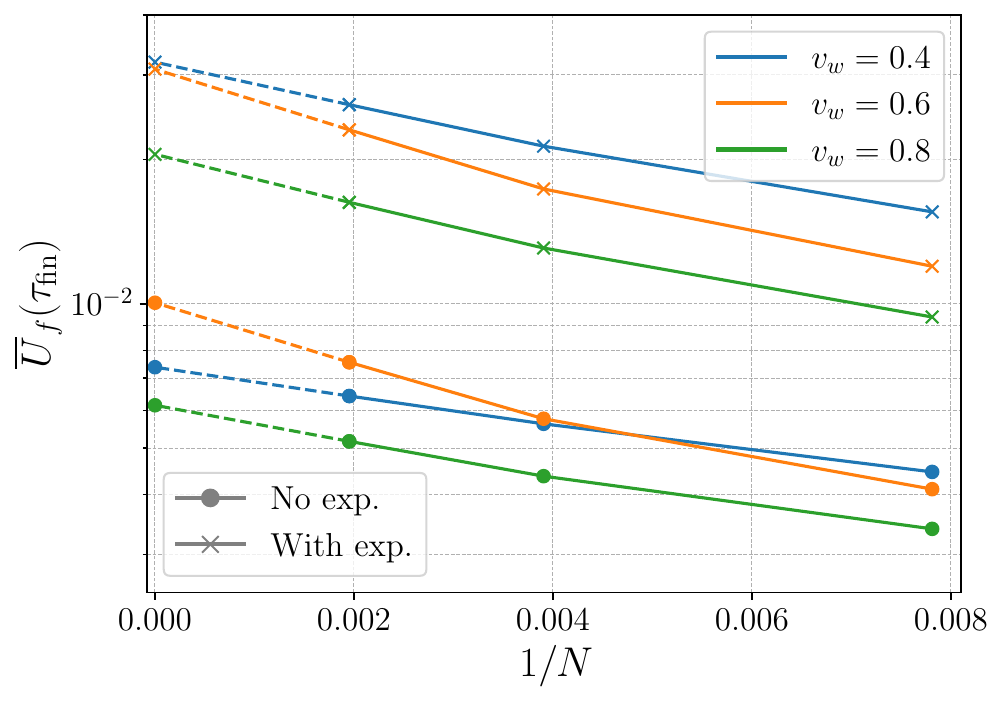}%
    \includegraphics[width=0.5\linewidth]{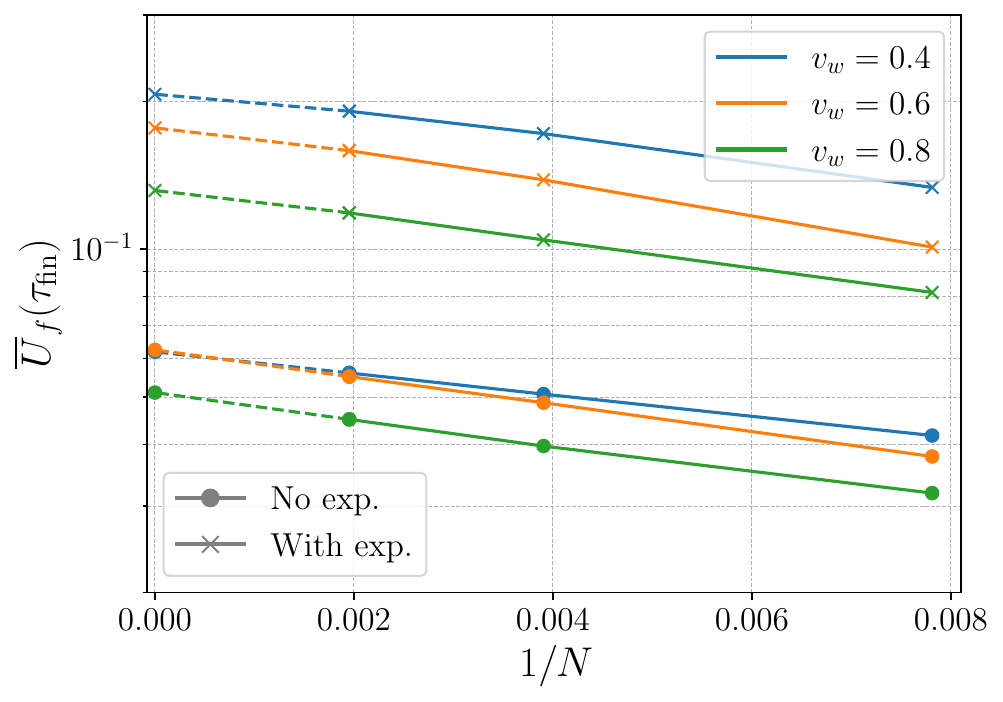}
    \caption{The RMS velocity $\overline{U}_f$ at $\tau_{\rm fin}$ in the simulations as the functions of $1/N$, where $N$ is the resolution, varying from $N=\{128, 256, 512\}$. The left panel represent results for a transition with weak strength ($\alpha_0=0.005$) and the right panel stand for intermediate transitions ($\alpha_0=0.05$). The dashed lines mark the result from linear extrapolation to $N\rightarrow \infty$. Stars in the right panel represent the estimated $\tau_{\rm fin}$ when considering cosmic expansion.}
    \label{fig:conv}
\end{figure}

To assess the numerical robustness of our simulations, we perform simulations at three different resolutions, $N=\{128,256,512\}$, and plot the relation between the maximum $\bar{U}_f$ and $1/N$, with extrapolations to $N\rightarrow \infty$. We find that the extrapolated values are typically $7\%$--$10\%$ higher than those obtained at $N=512$. Since $\Omega_{\rm GW} \propto \overline{U}_f^4$, this may lead to an underestimation of the GW content in simulations with this finite resolution. This is typical of Higgsless simulations and may arise from numerical dissipation associated with the presence of moving, infinitesimally thin walls. Such deviations may be more severe for the hybrid mode (around $v_w=0.6$) in weak transitions. However, since the expanding and non-expanding scenarios exhibit similar levels of underestimation in the GW content, our estimate of the ratio of the GW contents is robust.

\subsection{Fitting parameters}
We use a double-broken power-law fitting template, as adopted in Ref.~\cite{Jinno:2022mie}, to fit the GW spectra:
\begin{align}
    \Omega^*(k) = \Omega_0 \times \frac{(k/k_0)^3}{1+(k/k_0)^2\left[1+(k/k_1)^4\right]} \times e^{-(k/k_e)^2}\,.
\end{align}
The fitting parameters for the non-expanding and expanding cases are given in Tabs.~\ref{tab:fit-noexp} and~\ref{tab:fit-exp}, respectively.

\begin{table}[h]
\centering
\begin{minipage}{0.48\textwidth}
\caption{Fitting parameters in a non-expanding scenario.}
\centering
\label{tab:fit-noexp}
\begin{tabular}{ccccccc}
\toprule
$\beta/H_n$ & $v_w$ & strength & $\Omega_0$ & $k_0$ & $k_1$ & $k_e$ \\
\midrule
\midrule
15 & 0.3 & weak  & $1.14\times 10^{-11}$ & 29.72 & 43.53 & 408.32 \\
 &  & inter & $1.47\times 10^{-7}$  & 34.51 & 38.44 & 406.05 \\
 & 0.6 & weak  & $3.94\times 10^{-12}$ & 7.63  & 122.34 & 402.98 \\
 &  & inter & $1.90\times 10^{-8}$  & 7.27  & 107.64 & 428.08 \\
 & 0.9 & weak  & $8.78\times 10^{-12}$ & 15.98 & 29.21 & 303.01 \\
 &  & inter & $4.11\times 10^{-8}$  & 12.73 & 33.14 & 311.64 \\
\midrule
\midrule
20 & 0.3 & weak  & $7.18\times 10^{-12}$ & 38.96 & 62.92  & 445.37 \\
 &  & inter & $1.97\times 10^{-7}$  & 66.85 & 48.71  & 464.34 \\
 & 0.6 & weak  & $3.01\times 10^{-12}$ & 11.20 & 152.29 & 553.34 \\
 &  & inter & $1.38\times 10^{-8}$  & 10.31 & 130.52 & 640.01 \\
 & 0.9 & weak  & $2.53\times 10^{-11}$ & 39.04 & 30.56  & 398.91 \\
 &  & inter & $7.97\times 10^{-8}$  & 27.76 & 37.09  & 411.70 \\
\midrule
\midrule
25 & 0.3 & weak  & $6.94\times 10^{-12}$ & 56.01 & 79.27  & 500.19 \\
 &  & inter & $7.77\times 10^{-8}$  & 73.22 & 66.80  & 553.72 \\
 & 0.6 & weak  & $2.62\times 10^{-12}$ & 14.78 & 175.56 & 716.18 \\
 &  & inter & $8.68\times 10^{-9}$  & 11.77 & 152.10 & 1039.73 \\
 & 0.9 & weak  & $8.48\times 10^{-12}$ & 34.23 & 44.16  & 424.72 \\
 &  & inter & $3.49\times 10^{-8}$  & 26.31 & 50.62  & 446.23 \\
\bottomrule
\end{tabular}%
\end{minipage}
\hfill
\begin{minipage}{0.48\textwidth}
\centering
\caption{Fitting parameters in an expanding scenario.}
\label{tab:fit-exp}
\begin{tabular}{ccccccc}
\toprule
$\beta/H_n$ & $v_w$ & strength & $\Omega_0$ & $k_0$ & $k_1$ & $k_e$ \\
\midrule
\midrule
15 & 0.3 & weak  & $5.59\times 10^{-9}$ & 90.25 & 60.33 & 432.32 \\
 &  & inter & $1.04\times 10^{-7}$  & 20.13 & 93.19 & 555.40 \\
 & 0.6 & weak  & $4.39\times 10^{-11}$ & 4.11  & 163.81 & 397.48 \\
 &  & inter & $1.95\times 10^{-7}$  & 8.58  & 144.56 & 414.05 \\
 & 0.9 & weak  & $1.86\times 10^{-10}$ & 21.84 & 49.79 & 333.14 \\
 &  & inter & $3.28\times 10^{-7}$  & 13.47 & 61.84 & 365.87 \\
\midrule
\midrule
20 & 0.3 & weak  & $1.61\times 10^{-9}$ & 104.29 & 82.98  & 486.45 \\
 &  & inter & $2.40\times 10^{-8}$  & 23.12 & 122.63 & 716.85 \\
 & 0.6 & weak  & $3.26\times 10^{-11}$ & 8.51  & 197.02 & 551.88 \\
 &  & inter & $7.51\times 10^{-8}$  & 10.12 & 191.99 & 542.18 \\
 & 0.9 & weak  & $8.09\times 10^{-11}$ & 26.17 & 61.17  & 434.35 \\
 &  & inter & $1.33\times 10^{-7}$  & 14.80 & 76.99  & 476.28 \\
\midrule
\midrule
25 & 0.3 & weak  & $5.18\times 10^{-8}$ & 564.75 & 50.12  & 573.24 \\
 &  & inter & $1.03\times 10^{-8}$  & 28.10  & 147.19 & 877.03 \\
 & 0.6 & weak  & $3.38\times 10^{-11}$ & 17.43  & 214.85 & 761.01 \\
 &  & inter & $4.48\times 10^{-8}$  & 13.87  & 223.22 & 759.53 \\
 & 0.9 & weak  & $4.59\times 10^{-11}$ & 31.32  & 73.61  & 480.74 \\
 &  & inter & $9.82\times 10^{-8}$  & 20.17  & 88.44  & 552.83 \\
\bottomrule
\end{tabular}%
\end{minipage}
\end{table}

\subsection{Present-day GW spectrum}
To obtain the present-day GW spectrum $\Omega_{\rm GW}(f)$, one can multiply $\Omega_{\rm GW}^*(k)$ by the factor
\begin{equation}
    F_{\rm GW} = \Omega_{\gamma, 0}\left(\frac{g_{s0}}{g_{s*}}\right)^{4/3}\frac{g_*}{g_0}.
\end{equation}
Here, $\Omega_{\gamma, 0}$ is the energy density fraction of photons today, $g_s$ is the number of entropic degrees of freedom, and $g$ is the effective number of relativistic degrees of freedom. 
The subscripts $*$ and $0$ represent the values at the time of production and at present, respectively.
Adopting the Hubble parameter today $H_0 = 67.4\pm0.5\mathrm{km}~\mathrm{s}^{-1}\mathrm{Mpc}^{-1}$ and the current photon temperature $T_{\gamma,0}=2.72548\pm0.00057\mathrm{K}$, setting $g_{s*}=g_*$ for $T>0.1\mathrm{MeV}$, and taking $g_0=2$, $g_{s0}=3.91$, one obtains
\begin{equation}
    F_{\rm GW} = (3.57\pm0.05)\times10^{-5}\left(\frac{100}{g_*}\right)^{1/3}\,,
\end{equation}
with the redshifted frequencies given by
\begin{equation}
    f=2.6\times10^{-6}\mathrm{Hz}\frac{k}{H_n}\frac{T_n}{100\mathrm{GeV}}\left(\frac{g_*}{100}\right)^{1/6}\,.
\end{equation}

\end{document}